\documentclass[11pt,a4paper]{article}
\pdfoutput=1
\usepackage{jcappub}
\usepackage{bm}
\usepackage{upgreek}
\newcommand{\Mpl}{M_{\rm pl}}

\newcommand{\be}{\begin{equation}}
\newcommand{\ee}{\end{equation}}
\newcommand{\bea}{\begin{eqnarray}}
\newcommand{\eea}{\end{eqnarray}}
\newcommand{\eref}{\eqref}

\newcommand{\p}{{\bf p}}

\newcommand{\di}{ {\rm d}}

\graphicspath{{./}}
\notoc

\begin{document}

\title{Transporting non-Gaussianity from sub to super-horizon scales}

\author{David J. Mulryne}

\affiliation{Astronomy Unit, School of Physics and Astronomy, Queen Mary University of London, London, E1 4NS, UK}

\emailAdd{d.mulryne@qmul.ac.uk}

\abstract{We extend the `moment transport method'  for calculating the statistics of 
inflationary perturbations to the quantum phase of evolution on 
sub-horizon scales. The quantum transport equations form a set of coupled 
ordinary differential equations for the evolution of 
quantum correlation functions during inflation, which are valid on sub- and super-horizon scales, 
and reduce to the known 
classical transport equations after horizon crossing. The classical and quantum equations follow directly 
from the field equations of cosmological perturbation theory.
In this paper, we focus on how the evolution 
equations arise, and explore how transport methods relate to other approaches, and in particular  
how formal integral solutions to the transport equations connect to those of the In-In formalism.
}


\maketitle
\newpage

\section{Introduction}

The non-Gaussian statistics of the curvature perturbation, $\zeta$, 
produced during inflation, have become a  
key test of inflationary models.  Methods to calculate these statistics are now 
well established. Typically the calculation is split into two parts, 
one for the quantum `sub-horizon' regime, 
and one for the classical `super-horizon' regime. It is common to perform the 
former part of the calculation 
using standard techniques of quantum field theory. First the interaction picture is employed, then 
initial conditions are provided by assuming linear Fourier 
modes whose wavelength is far smaller than the cosmological horizon are in the Bunch-Davis vacuum
state, and finally 
same-time correlation functions are calculated in the interacting vacuum.
This calculation is often described as the In-In formalism 
applied to inflationary cosmology (see for example Refs.~\cite{Maldacena:2002vr, Seery:2005wm, Chen:2006nt, Seery:2005gb,Elliston:2012ab}
for calculations of the inflationary bispectrum). 
It is usual to evaluate these quantum 
correlation functions shortly after the Fourier modes of interest cross 
the cosmological horizon, where for 
single field models $\zeta$, and hence the correlation functions of 
$\zeta$, become constant \cite{Rigopoulos:2003ak,Lyth:2004gb}. 

In multiple-field models, a complication is that $\zeta$ can continue to evolve. 
In principle there is no barrier to evaluating the correlation functions at 
later times using the In-In formalism, thus accounting for this evolution. But in practice this 
becomes extremely difficult analytically 
 (see Ref.~\cite{Dias:2012qy} for recent progress in that direction), 
and moreover the integrals which one is required to evaluate are not particularly 
well suited to numerical integration. The usual procedure, therefore, is to truncate the full 
quantum calculation at horizon crossing, to interpret the correlation functions as 
classical statistical quantities, and to employ classical methods to follow 
the statistics from that point on \cite{Lyth:2005fi,Seery:2005wm}. This is valid since the dynamics is expected to 
be very close to classical in this regime. 

Despite the fairly mature state of existing techniques, it is still useful to explore 
new methods to calculate the statistics of inflationary perturbations for several reasons. 
First, with the 
imminent expected influx of data,
computationally efficient algorithms are urgently needed to precisely compare the predictions of 
inflationary models 
with observations. Secondly, in this connection, it would be useful to have a computationally efficient method for 
which a splitting into sub- and super-horizon regimes was not required. This will be invaluable for 
many models, such as those in which the inflationary trajectory bends during horizon crossing, where such a splitting 
will not yield accurate results. 
Finally, there is considerable interest in understanding the 
exact relation between 
sub-horizon and super-horizon methods. For example, in order to calculate the error incurred by 
using purely the latter as is commonly done in studies of multiple field inflation. One approach to 
understanding this link is to consider how classical equations 
emerge from the quantum system via the dynamical renormalisation group flow \cite{Dias:2012qy}. In this 
paper we pursue a complementary route.

The most
widely applied super-horizon approach is the 
$\delta N$ formalism which utilises the separate universe approximation \cite{Lyth:1984gv,Bardeen:1983qw, Wands:2000dp,Lyth:2005fi,Malik:2005cy}. 
Recently another technique called the 
`moment transport method' has also been extensively explored \cite{Mulryne:2009kh,Mulryne:2010rp,Dias:2011xy,Elliston:2012ab} 
(also see \cite{Yokoyama:2007uu,Yokoyama:2007dw,Yokoyama:2008by} for related 
earlier work). One can show that objects which provide formal solutions to the transport system (see \S \ref{transportProperties}) 
are equivalent to the coefficients of a 
$\delta N$ style Taylor expansion \cite{Seery:2012vj,Anderson:2012em}, and hence the methods are formally equivalent. The transport 
approach, however, has the computational advantage of providing a 
simple set of coupled evolution equations 
for the classical two-point, three-point, and higher, correlation functions, or for the Taylor expansion coefficients \cite{Anderson:2012em}. 
Such equations provide important analytic insights into the evolution of the statistics \cite{Elliston:2011dr,Elliston:2011et,Seery:2012vj}, and
are amenable to a straightforward numerical implementation (see, for example, Refs.~\cite{Frazer:2011br,Dias:2012nf} for studies utilising 
numerical implementations of the transport equations).

In this paper, our aim is to show how the transport equations of motion 
can be extended to sub-horizon scales 
to provide a set of 
coupled ordinary differential equations (ODEs) 
for quantum correlation functions\footnote{Similar equations are used in other settings (see \cite{Bojowald:2005cw} and references therein), 
and provided inspiration for looking at this approach in the inflationary setting.}, which are valid in 
both sub- and super-horizon regimes, and tend to the classical equations after horizon crossing. 
This offers an alternative route to calculating the statistics of $\zeta$. Moreover, since differential 
equation formulations of a problem are often easier to implement numerically than integral formulations, 
the quantum transport equations will likely also be of  benefit for numerical implementations of the 
full quantum and classical system. Though we defer such a implementation to future work \cite{DFMS}.

The layout of the present work is as follows. In \S \ref{classicalTransport} we review the transport 
approach to classical 
statistics, and discuss how the transport equations look for equations of motion in which 
spatial gradient terms have not been discarded, as they typically are on super-horizon scales where 
transport methods have previously been applied. In \S \ref{transportProperties} we review some analytical 
properties of the moment transport equations and related transport equations. 
Then in \S \ref{quantumTransport} we show how 
the formalism can be further extended to regimes in which the field perturbations 
behave as quantum objects, as they do 
on sub-horizon scales.  We also show how formal integral 
solutions to these quantum equations are connected to the integral expressions of the In-In formalism. 
In \S \ref{singleField}, we explore these results using an explicit example of canonical single field inflation.
Finally,  in \S \ref{multiField}, we show that 
the transport expressions for the two-point function in multiple field settings recover 
known methods originally described by Salopeck {\it et al.} \cite{Salopek:1988qh}. 
We conclude in \S \ref{conclusions} and discuss future directions. 

\section{Transporting classical statistics}
\label{classicalTransport}

The simplest way to derive the transport equations is to begin with  
the evolution equations for cosmological perturbations themselves\footnote{One can also 
begin with the partial differential equation for the joint probability distribution of inflationary 
perturbations and decompose this equation into ordinary differential equations for each moment 
or cumulant \cite{Mulryne:2009kh,Mulryne:2010rp}.}.
In Fourier space, we can utilise a form of DeWitt notation to write these perturbation equations in a highly compact form \cite{Seery:2012vj,Anderson:2012em}. 
In this notation, $x_{\alpha'}$ represents the Fourier modes of all fields and field velocity perturbations 
defined on flat hypersurfaces and evaluated at some time $t$\footnote{Or any other complete 
set of perturbations which completely parametrise the perturbed spacetime, for example the curvature and isocurvature modes, can be used.
In this paper we will 
always consider  fields and field velocity perturbations.}.  
A primed index, for example $\alpha'$, on any object indicates a field/field velocity label $\alpha$ and a momentum label
$\mathbf{k}_{\upalpha}$, such that for example 
\be
x_{\alpha'} = x_\alpha(\mathbf{k}_{\mathsf{\upalpha}})\,.
\ee
$\alpha$ therefore runs from $1$ to $2 n$ where $n$ is the total number of fields, 
and for odd $\alpha$, $x_{\alpha'} = \delta \phi_{ (\alpha+1)'/2}$, while for even $\alpha$,  $x_{\alpha'}= \delta \dot \phi_{\alpha'/2}$.
Repeated 
primed indices imply that the fields and field velocities 
are summed over {\it and} an 
integration is performed over the suppressed Fourier space dependence, 
with measure $\di^3 k_\upalpha / (2 \pi)^3$. Repeated 
non-primed indices indicate just summation over the field label, 
leaving the Fourier mode unchanged as the sum runs through each value of $\alpha$. Below we will see an explicit example of this. 

In 
this notation, perturbations obey the equations of motion 
\be
\label{perts101}
\frac{\di  x_{\alpha^\prime}}{\di t} =u_{\alpha' \beta'}  x_{\beta'} + \frac{1}{2} u_{\alpha' \beta' \gamma'}  \left (x_{\beta'} x_{\gamma'} - \langle  x_{\beta'} x_{\gamma'} \rangle \right )\,,
\ee
where the $u$ coefficients 
can be derived from cosmological perturbation theory (for a recent review see Ref.~\cite{Malik:2008im}), 
or on super-horizon scales 
from alternative methods utilising the separate universe approach \cite{Yokoyama:2007uu,Yokoyama:2007dw,Mulryne:2009kh,Mulryne:2010rp,Anderson:2012em}.
The form of $u_{\alpha' \beta'}$ is
\be
\label{u1}
 u_{\alpha' \beta'} = (2 \pi)^{3} u_{\alpha \beta}(k_\upalpha) \delta(\mathbf{k_\upalpha} - \mathbf{k_\upbeta}) \,,
\ee 
where we sometimes describe $u_{\alpha \beta}(k_\upalpha)$ (distinguished by the unprimed indices) as the kernel of $u_{\alpha' \beta'}$, and, 
for example,
\begin{eqnarray}
u_{\alpha'\beta'}  x_{\beta'} &=& \int \left. \di^3 k_\upbeta  u_{\alpha \beta}(k_\upalpha) \delta(\mathbf{k_\upalpha} - \mathbf{k_\upbeta})  x_\beta(\mathbf{k_\upbeta}) \right.\,,
\nonumber\\  
&=&   u_{\alpha \beta}(k_\upalpha)  x_\beta(\mathbf{k_\upalpha}) \,,
\end{eqnarray} 
where on the right hand side of the final expression the only sum is over the 
repeated index $\beta$, and $k_\upalpha$ is simply one 
particular Fourier mode, which we could have equally well labelled $k$, $k_1$, $k_2$, $k'$ or $k''$ etc. as is common in the literature.
Likewise, the form of $u_{\alpha' \beta' \gamma'}$ is  
\be
\label{u2}
u_{\alpha' \beta' \gamma'} = (2 \pi)^{3} u_{\alpha \beta \gamma} (k_\upalpha, k_\upbeta, k_\upgamma) \delta(\mathbf{k}_\upalpha - \mathbf{k_\upbeta} - \mathbf{k_\upgamma}) \,,
\ee 
and for example,
\begin{eqnarray}
u_{\alpha'\beta' \gamma'}  x_{\beta'} x_{\gamma'} &=& \frac{1}{(2 \pi)^3}\int \left. \int \left. \di^3 k_\upbeta \di^3 k_\upgamma u_{\alpha \beta \gamma}(k_\upalpha, k_\upbeta, k_\upgamma) \delta(\mathbf{k}_\upalpha - \mathbf{k}_\upbeta - \mathbf{k}_\upgamma)  x_\beta(\mathbf{k}_\upbeta) x_\gamma(\mathbf{k}_\upgamma)\right. \right.\,,
\nonumber\\  
&=&   \frac{1}{(2 \pi)^3}\int \left. \di^3 k_\upbeta u_{\alpha \beta \gamma}(k_\upalpha, k_\upbeta, k_\upalpha-k_\upbeta)  x_\beta(\mathbf{k}_\upbeta) x_\gamma(\mathbf{k}_\upalpha-\mathbf{k}_\upbeta)\right. \,.
\end{eqnarray}
The form of $u$ coefficients given in Eqs.~\eref{u1} and \eref{u2} are such that after substituting into Eq.~\eref{perts101}, 
the various terms take the expected form, which follows from Fourier transforming the original real space equation of motion. In 
particular the second term on the RHS of Eq.~\eref{perts101} involves a convolution over all scales. 
In general the kernels of the $u$ coefficients are functions of background quantities and can also carry 
Fourier mode dependence as indicated above. 
For super-horizon regimes they become independent of any scale dependence.
In \S \ref{singleField} we will see explicit examples of the $u$ coefficients which 
arise from cosmological perturbation theory.

The notation involving primed indices we have introduced in this section may seem unnecessary. However, we can already see
from Eq.~\eref{perts101} that it allows complicated expressions, involving for example convolutions, to be written in 
a compact form utilising an extended summation convention. This will become increasingly advantageous for manipulations as we proceed.

\subsection{Evolution of statistics}
In order to turn the evolution equations for field fluctuations into 
evolution equations for their statistics, we employ the simple principle
that expectation values obey the equation  
\be
\label{stats101}
 \frac{d \langle A \rangle}{d t} = \left \langle \frac{d A}{d t} \right   \rangle\,,
\ee
where A is any function of perturbations which has no explicit time 
dependence. 
As discussed in more detail in Refs.~\cite{Seery:2012vj,Anderson:2012em}, this follows by considering the fact that 
an evolving classical distribution of 
perturbations must conserve probability.

By defining the two and three-point correlations functions as 
\be
\Sigma_{\alpha' \beta'} \equiv  \left \langle x_{\alpha'}  x_{\beta'} \right \rangle  \,,  
\ee
and 
\be
 \alpha_{\alpha' \beta' \gamma'} \equiv \left \langle x_{\alpha'} x_{\beta'} x_{\gamma'} \right \rangle \,, 
\ee
respectively, identifying $\langle A \rangle$ with $\Sigma_{\alpha' \beta'}$ and then with $\alpha_{\alpha' \beta' \gamma'} $,
and finally utilising Eqs.~\eref{stats101} and \eref{perts101} and the chain rule, one finds
a set of coupled equations 
\begin{eqnarray}
\label{tbasic}
\frac{\di\Sigma_{\alpha^{\prime}\beta^{\prime}}}{\di t}
	&=&
		u_{\alpha^{\prime}\gamma^{\prime}}
		\Sigma_{\gamma^{\prime}\beta^{\prime}}
		+
		u_{\beta^{\prime}\gamma^{\prime}}
		\Sigma_{\gamma^{\prime}\alpha^{\prime}}+\dots\,, \nonumber \\
\frac{\di\alpha_{\alpha^{\prime}\beta^{\prime}\gamma^{\prime}}}{\di t}
	&=&
		u_{\alpha^{\prime}\lambda^{\prime}}
		\alpha_{\lambda^{\prime}\beta^{\prime}\gamma^{\prime}}
		+
		u_{\alpha^{\prime}\lambda^{\prime}\mu^{\prime}}
		\Sigma_{\lambda^{\prime}\beta^{\prime}}
		\Sigma_{\mu^{\prime}\gamma^{\prime}}
		+ 		\left (		\alpha'\rightarrow\beta'\rightarrow\gamma' \right)~+\dots \,,
\end{eqnarray}
where the dots indicate we have truncated the equations to leading order, and the arrow that there are 
two additional terms formed by cyclic permutations. Here we have considered 
only the two and three-point correlation functions, and for simplicity 
we will restrict ourselves to these 
throughout this paper. Similar expressions exist for the four-point function \cite{Anderson:2012em} and can be calculated 
for the higher $n$-point functions in the same manner. 
These equations allow known classical statistics at some point in time to be propagated 
forward to a later point 
at which we wish to evaluate them (for example between horizon crossing and the end of inflation). 

\subsection{Equations for correlations are simpler than equations for perturbations}

The above equations are complicated because they involve integrations 
over $k$ space, which follow from the convolutions which 
naturally appear in Fourier space perturbation equations beyond linear order.  
Correlation functions in Fourier space, however, are always accompanied with delta functions, and at any given time, $t$, 
we have
\be 
\label{tsimple}
\Sigma_{\alpha' \beta'} = (2\pi)^3 \Sigma_{\alpha \beta}(k_\upalpha) \delta(\mathbf{k}_\upalpha + \mathbf{k}_\upbeta)\,,
\ee 
and
\be
\alpha_{\alpha' \beta' \gamma'} = (2\pi)^3 \alpha_{\alpha \beta \gamma}(k_\upalpha, k_\upbeta, k_\upgamma)  \delta(\mathbf{k}_\upalpha + \mathbf{k}_\upbeta+ \mathbf{k}_\upgamma)\,, 
\ee
where we refer to $\Sigma_{\alpha \beta}(k_\upalpha)$ and $\alpha_{\alpha \beta \gamma}(k_\upalpha, k_\upbeta, k_\upgamma)$ as the kernel of 
the two- and three-point functions respectively\footnote{Note that for convenience, the notation is slightly different to Ref.~\cite{Anderson:2012em}.}.
We find that
\begin{eqnarray}
\frac{\di\Sigma_{\alpha \beta}(k_\upalpha)}{\di t}
	&=&
		u_{\alpha \gamma}(k_\upalpha) \Sigma_{\gamma \beta}(k_\upalpha)
		+
		u_{\beta \gamma}(k_\upalpha) \Sigma_{\gamma \alpha}(k_\upalpha)
		+\dots\,, \nonumber\\
\frac{\di\alpha_{\alpha \beta \gamma}(k_\upalpha, k_\upbeta, k_\upgamma)}{\di t}
	&=&
		u_{\alpha \lambda}(k_\upalpha)
		\alpha_{\lambda \beta \gamma}(k_\upalpha, k_\upbeta, k_\upgamma) + u_{\alpha \lambda \mu }(k_\upalpha, k_\upbeta, k_\upgamma)
		\Sigma_{\lambda \beta}(k_\upbeta)
		\Sigma_{\mu \gamma}(k_\upgamma)
		\nonumber\\&+& \left(
\alpha\rightarrow \beta \rightarrow \gamma \right ) + \dots\,,
\label{tsimple2}
\end{eqnarray}
where once again we give only the leading order terms in each equation. In the final expression it should be 
understood that 
moving the position of free indices when performing the cyclic permutations, $\alpha \rightarrow \beta \rightarrow \gamma$,
also moves the 
positions of the associated Fourier modes $k_\upalpha \rightarrow k_\upbeta \rightarrow k_\upgamma$.
An explicit example of a system of this type is given in \S \ref{singleField}.
These equations 
exhibit the feature that one does not need to 
perform a convolution over all scales in order to calculate the evolution of 
$\alpha_{\alpha \beta \gamma}(k_\upalpha,k_\upbeta,k_\upgamma)$, as 
one does in Eq.~\eref{perts101} for the evolution of $x_{\alpha}(\mathbf{k}_\upalpha$). In this sense 
Eqs.~\eref{tsimple2} are simpler than Eq.~\eref{perts101}.

\section{Other properties of the transport system}
\label{transportProperties}

The transport system, given by Eqs.~\eref{tbasic}, has many attractive properties. 
In particular, one can introduce 
objects, closely related 
to the correlations themselves, which allow formal analytic solutions of the transport hierarchy, 
using path-ordered exponentials (see Refs.~\cite{Seery:2012vj,Anderson:2012em}). 
Here we simply quote the results, and point out how they appear in a scale dependent setting.

The solutions of Eqs.~\eref{tbasic} from some initial time $t_0$ to some later
time $t$ are given by 
\begin{eqnarray}
\label{sigmaSol}
\Sigma_{\alpha' \beta'}&=&\Gamma_{\alpha' i'} \Gamma_{\beta' j'}\Sigma_{i' j'}\,,  \\
\alpha_{\alpha' \beta' \gamma'}=\Gamma_{\alpha' i'} \Gamma_{\beta' j'} \Gamma_{\gamma' k'}\alpha_{i' j' k'} 
					&+& \left [\Gamma_{\alpha' i' j'} \Gamma_{\beta' k'} \Gamma_{\gamma' l'} \Sigma_{i'k'} \Sigma_{j' l'}\right.\, \nonumber \\
					&+&  \left. \left( \alpha' \rightarrow \beta' \rightarrow \gamma' \right )\right]\,,
\label{alphaSol}
\end{eqnarray}
where we have adopted further compact notation according to which a Greek subscript indicates evaluation 
at the later time $t$ as before, and a Roman index indicates evaluation at the initial time $t_0$, and where we have  
presented only the leading order terms in these solutions.
Mixed index objects have dependence on both 
times. Here we have introduced a new function $\Gamma_{\alpha' i'}$ which is the propagator of the system. $\Gamma_{\alpha' i'}$
satisfies its own transport equation 
\be
\label{gamma1Evolve}
	\frac{\di \Gamma_{\alpha' i' }}{\di t}
	= u_{\alpha' \beta'} \Gamma_{\beta' i'} \,,
\ee
and we have also introduced $\Gamma_{\alpha' i' j'}$ which obeys the transport equation 
\be
\label{gamma2Evolve}
	\frac{\di \Gamma_{\alpha' i' j'}}{\di t}
	= u_{\alpha' \beta'} \Gamma_{\beta' i' j'} +
	u_{\alpha' \beta' \gamma'} \Gamma_{\beta' i'} \Gamma_{\gamma' j'}  \,.
\ee
The initial conditions are $\Gamma^{\rm init}_{\alpha' i'}=\delta_{\alpha' i'}$  and $\Gamma^{\rm init}_{\alpha' i' j'}=0$, which follow from 
their definition in Eqs.~\eref{sigmaSol} and \eref{alphaSol}.
These objects can be solved in terms of a path ordered exponential to find
\begin{eqnarray}
	\Gamma_{\alpha' i'} &=& 
	{\cal P} \exp \left (
		\int_{t_0}^{t} \di t' \; u_{\tilde \alpha' \tilde\gamma'}(t') 
	\right) \delta_{ \gamma' i'} \,,\nonumber\,\\
	\Gamma_{\alpha' i' j'} &=& \int_{t_0}^{t} \di t'  \Gamma_{\alpha' \tilde \mu'} u_{\tilde \mu' \tilde \nu' 
	\tilde \sigma' }(t') \Gamma_{\tilde \nu  i'} \Gamma_{\tilde \sigma' j'}\,, 
	\label{gammaSols}
\end{eqnarray}
where the indices marked with a tilde, indicate they are associated with the intermediate time in the integral $t'$. 

In terms of these solutions, Eq.~\eref{alphaSol} becomes
\begin{eqnarray}
\alpha_{\alpha' \beta' \gamma'} &=& \Gamma_{\alpha' i'} \Gamma_{\beta' j'} \Gamma_{\gamma' k'}\alpha_{i' j' k'} 
					+  \int_{t_0}^{t} \di t'  \left [\Gamma_{\alpha' \tilde \mu'} u_{\tilde \mu' \tilde \nu' 
	\tilde \sigma' } \Sigma_{\beta' \tilde \nu'} \Sigma_{\gamma' \tilde \sigma'} + \left( \alpha' \rightarrow \beta' \rightarrow \gamma' \right ) \right ],
\label{alphaSol2a}
\end{eqnarray}
where we have introduced mixed indexed two-point correlations, 
$\Sigma_{\alpha'  i'} = \Gamma_{\alpha' k'} \Sigma_{k' i'}$, which by construction satisfy the 
same equation of motion as $\Gamma_{\alpha' i'}$. 

Once again the expressions above can be simplified, since consistency 
implies that 
\be
\Gamma_{\alpha' i'} = (2 \pi)^3\Gamma_{\alpha i}(k_\upalpha) \delta(\mathbf{k}_\upalpha - \mathbf{k}_{\rm i}) \,,
\ee
 and 
\be
\Gamma_{\alpha' i' j'} = (2 \pi)^3\Gamma_{\alpha i j}(k_\upalpha, k_{\rm i}, k_{\rm j}) \delta(\mathbf{k}_\upalpha - \mathbf{k}_{\rm i}-\mathbf{k}_{\rm j})\,. 
\ee 
As in the previous sections, therefore, all integrations over Fourier modes can be explicitly 
performed in the equations of motion \eref{gamma1Evolve} and \eref{gamma2Evolve}, and in the 
integral solutions.  We find, for example,
\begin{eqnarray}
	\Gamma_{\alpha i}(k_\upalpha) &=& 
	{\cal P} \exp \left (
		\int_{t_0}^{t} \di t' \; u_{\tilde \alpha \tilde\gamma}(k_\upalpha,t') 
	\right) \delta_{\tilde \gamma i} \,,\nonumber\,\\
	\Gamma_{\alpha i j}(k_\upalpha, k_{\rm i}, k_{\rm j}) &=& \int_{t_0}^{t} \di t'  \Gamma_{\alpha \tilde \mu}(k_\upalpha) u_{\tilde \mu \tilde \nu 
	\tilde \sigma } (k_\upalpha, k_{\rm i}, k_{\rm j}) \Gamma_{\tilde \nu  i}(k_{\rm i}) \Gamma_{\tilde \sigma j}(k_{\rm j})\,, 
	\label{gammaSols2}
\end{eqnarray}
and
\begin{eqnarray}
\alpha_{\alpha \beta \gamma}&=&\Gamma_{\alpha i}(k_\upalpha) \Gamma_{\beta j}(k_\upbeta) \Gamma_{\gamma k}(k_\upgamma) \alpha_{i j k}(k_\upalpha, k_\upbeta, k_\upgamma) \nonumber \\
					&+& \left[ \int_{t_0}^{t} \di t'  \Gamma_{\alpha \tilde \mu}(k_\upalpha) u_{\tilde \mu \tilde \nu 
	\tilde \sigma }(k_\upalpha, k_\upbeta, k_\upgamma) \Sigma_{\beta \tilde \nu}(k_\upbeta) \Sigma_{\gamma \tilde \sigma}(k_\upgamma)\nonumber \right.\\
&+&  \left( \alpha \rightarrow \beta \rightarrow \gamma \right ) \Big] \,.
\label{alphaSol2b}
\end{eqnarray}
One can verify that direct differentiation of Eq.~\eref{alphaSol2b} gives the evolution equation \eref{tsimple2}.

\section{Quantum transportation}
\label{quantumTransport}

We now turn to quantum transport equations. Our aim is derive analogous equations and solutions to those above, but valid even 
when the quantum properties of the fluctuations are important. This will mean that if values of inflationary correlation 
functions are known at some initial time, even if that time is long before horizon crossing, they can be propagated forward to some later time, such 
as end of inflation. 

To do this we first promote 
$\delta \phi_\alpha(\mathbf{k_\alpha})$ to an operator $\widehat{\delta \phi_\alpha}(\mathbf{k_\alpha})$. This will be accompanied by a canonical 
momentum $ p_\alpha = \partial {\cal L} / \partial \delta \dot \phi_\alpha$, where ${\cal L}$ is 
the Lagrangian which 
leads to Eq.~\eref{perts101}. $\widehat{\delta \phi_\alpha}$ and $ \widehat{ p}_\alpha$ 
form a conjugate pair obeying the 
usual commutation relations 
$[\widehat{\delta \phi}_{\alpha'}, \widehat{ p}_{\beta'}] = i  \delta_{\alpha \beta} \delta(\mathbf{k_\alpha} - \mathbf{k_\beta})$, 
where we have adopted the 
same primed notation as above. 
We note that $p_{\alpha'}$ is not necessarily $\dot{\delta \phi_{\alpha'}}$, as it would be 
for a linear theory.

\subsection{Heisenberg picture and Ehrenfest's theorem}

At this point, working in the Heisenberg picture of quantum mechanics aids our discussion. 
In the Heisenberg picture, the operators are time dependent, while the 
states are independent of time.
Operators follow evolution equations of the form
\be
\frac{d \hat A}{d t} = -i [\hat A,\hat H]\,,
\ee
where $\hat A$ is some operator which has no explicit time dependence, and $\hat H$ 
is the Hamiltonian operator. 
As is well known, the explicit form 
of this equation for a given $\hat A$ is 
identical to the evolution equation for the 
classical variable $A$, but written in terms of operators. 
This is true only up to operator ordering, and a specific ordering for the Hamiltonian must be 
chosen. Here we assume that the Hamiltonian is fully symmetric. This is known as 
Weyl ordering. 
When we turn to correlation functions, we will see that if we take these to be Weyl ordered as well, then we only ever have 
to deal with real valued quantities. This is a choice which can be made for convenience. There is no loss 
of useful information if this choice is made since, 
ultimately, we evaluate the statistics on super-horizon regimes where they become classical and the ordering 
is irrelevant.

In this picture 
\be
\label{operatorPerts101}
\frac{d \widehat{ x}_{\alpha'}}{d t} =u_{\alpha' \beta'} \widehat{ x}_{\beta'} + \frac{1}{2} u_{\alpha' \beta' \gamma'} \left (\widehat{ x}_{\beta'}  \widehat{ x}_{\gamma'} - \langle \widehat{ x}_{\beta'}  \widehat{ x}_{\gamma'} \rangle \right ) ,
\ee
where $\widehat{ x}_\alpha$ are the operators associated with $\delta \phi_{(\alpha +1)/2}$ and $\delta \dot \phi_{\alpha/2}$ for odd and even 
$\alpha$ respectively, and  
$u_{\alpha' \beta' \gamma'}$ is symmetric in the last two indices. 
Now we wish to form evolution equations for the quantum correlation functions of the form 
\be
\Sigma_{\alpha' \beta'} = \frac{1}{2}\langle \widehat{ x}_{\alpha'} \widehat{ x}_{\beta'} \rangle\,,
\label{Squant1}
\ee
where for convenience we have used the same notation as for the classical object, but now 
it is understood that  $\Sigma_{\alpha' \beta'}$ is not in general symmetric in its two indices. 
We also have 
\be
\label{Aquant1}
\alpha_{\alpha' \beta' \gamma'} =  \langle \widehat{ x}_{\alpha'} \widehat{ x}_{\beta'} \widehat{ x}_{\gamma'} \rangle\,,
\ee
where we have  used the same notation as for the classical three-point function but now there is 
no longer symmetry in the ordering of indices. 
The evolution of 
any expectation value follows from Ehrenfest's theorem. In the Heisenberg picture, this amounts 
to Eq.~(\ref{stats101}), but with $A$ taken to be a quantum operator. This leads to 
\be
\label{Ehrenfest}
\frac{\di \langle  \hat A \rangle}{\di t} = \left \langle -i [\hat A,\hat H] \right \rangle\,,
\ee
which is the quantum equivalent of Eq.~\eref{stats101}.

At this point similar reasoning to that in the previous sections can be applied. Identifying $\langle \hat A \rangle$
with $\Sigma_{\alpha' \beta'}$ in Eq.~\eref{Squant1} and then with $\alpha_{\alpha' \beta' \gamma'}$ in Eq.~\eref{Aquant1}, 
considering Eq.~\eref{Ehrenfest} and \eref{operatorPerts101} and using Wick's theorem, we arrive at 
\begin{eqnarray}
\label{tqbasic}
\frac{\di\Sigma_{\alpha^{\prime}\beta^{\prime}}}{\di t}
	&=&
		u_{\alpha^{\prime}\gamma^{\prime}}
		\Sigma_{\gamma^{\prime}\beta^{\prime}}
		+
		u_{\beta^{\prime}\gamma^{\prime}}
		\Sigma_{\alpha^{\prime}\gamma^{\prime}} , \nonumber \\
\frac{\di\alpha_{\alpha^{\prime}\beta^{\prime}\gamma^{\prime}}}{\di t} 
	&=&
		u_{\alpha^{\prime}\lambda^{\prime}}
		\alpha_{\lambda^{\prime}\beta^{\prime}\gamma^{\prime}}
		+
		u_{\beta^{\prime}\lambda^{\prime}}
		\alpha_{\alpha' \lambda^{\prime}\gamma^{\prime}}
		+		u_{\gamma^{\prime}\lambda^{\prime}}
		\alpha_{\alpha' \beta '\lambda^{\prime}}\,\nonumber \\
		&+& u_{\alpha^{\prime}\lambda^{\prime}\mu^{\prime}}
		\Sigma_{\lambda^{\prime}\beta^{\prime}}
		\Sigma_{\mu^{\prime}\gamma^{\prime}} + 
		u_{\beta^{\prime}\lambda^{\prime}\mu^{\prime}}
		\Sigma_{\alpha' \lambda^{\prime}}
		\Sigma_{\mu^{\prime}\gamma^{\prime}} +
		u_{\gamma^{\prime}\lambda^{\prime}\mu^{\prime}}
		\Sigma_{\alpha' \lambda^{\prime}}
		\Sigma_{\beta ' \mu^{\prime}}  \,.
\end{eqnarray}
In deriving this equation we have used the chain rule in this 
quantum setting.  For example,  $[\hat x_{\alpha'} \hat x_{\beta'}, \hat H ] =[\hat x_{\alpha'}, \hat H ] \hat x_{\beta'} +   \hat x_{\alpha'} [\hat x_{\beta'} , \hat H ]$.
As before, since the quantum two and three-point functions are still accompanied by delta functions, 
the Fourier space integrals can be performed explicitly leading to analogous equations 
to Eqs.~\eref{tsimple} and \eref{tsimple2}. Indeed the equations are identical to those earlier equations 
except that the indices are ordered in the same 
way as in the equations \eref{tqbasic} above.
Moreover, formal solutions are possible. In this case, since ordering is again important, we have 
\begin{eqnarray}
\Sigma_{\alpha' \beta'} &=& \Gamma_{\alpha' i'} \Gamma_{\beta' j'} \Sigma_{i' j'}\, \nonumber \\
\alpha_{\alpha' \beta' \gamma'}&=&\Gamma_{\alpha' i'} \Gamma_{\beta' j'} \Gamma_{\gamma' k'}\alpha_{i' j' k'} 
					+ \Gamma_{\alpha' i' j'} \Gamma_{\beta' k'} \Gamma_{\gamma' l'} \Sigma_{i' k'} \Sigma_{ j' l'} \nonumber \\
					&+&   \Gamma_{\beta' i' j'} \Gamma_{\alpha' k'} \Gamma_{\gamma' l'} \Sigma_{k' i'} \Sigma_{j' l'}
					+  \Gamma_{\gamma' i' j'} \Gamma_{\alpha' k'} \Gamma_{\beta' l'} \Sigma_{k' i'} \Sigma_{l' j'}\,.\
\label{alphaSolq}
\end{eqnarray}
where the $\Gamma$ functions are identical to the classical case and follow identical evolution equations. Therefore,
the same formal integral solutions for the $\Gamma$ functions hold, and moreover we find
\begin{eqnarray}
\alpha_{\alpha \beta \gamma}(k_\upalpha,k_\upbeta,k_\upgamma)&=&\Gamma_{\alpha i}(k_\upalpha) \Gamma_{\beta j}(k_\upbeta) \Gamma_{\gamma k}(k_\upgamma) \alpha_{i j k}(k_\upalpha, k_\upbeta, k_\upgamma)
		\nonumber \\
					&+& \int_{t_0}^{t} \di t' \left[  \Gamma_{\alpha \tilde \mu}(k_\upalpha) u_{\tilde \mu \tilde \nu 
	\tilde \sigma }(k_\upalpha, k_\upbeta, k_\upgamma) \Sigma_{\tilde \nu \beta}(k_\upbeta) \Sigma_{\tilde \sigma \gamma }(k_\upgamma) \right.\nonumber \\
      &+& \left.\Gamma_{\beta \tilde \mu}(k_\upbeta) u_{\tilde \mu \tilde \nu 
	\tilde \sigma }(k_\upbeta, k_\alpha, k_\upgamma) \Sigma_{\alpha \tilde \nu}(k_\upalpha) \Sigma_{\tilde \sigma \gamma}(k_\upgamma) \right.\nonumber \\             
      &+& \left. \Gamma_{\gamma \tilde \mu}(k_\upgamma) u_{\tilde \mu \tilde \nu 
	\tilde \sigma }(k_\upgamma, k_\upalpha, k_\upbeta) \Sigma_{\alpha \tilde \nu }(k_\upalpha)\Sigma_{\beta \tilde \sigma}(k_\upbeta) 
\right  ] 
\,.
\label{alphaSol2q}
\end{eqnarray}
One can verify that direct differentiation of Eq.~\eref{alphaSol2q} gives the correct evolution equation.

\subsection{Connection to the In-In formalism}

In the In-In formalism, the integral expression for a correlation function (for example a three-point function with an 
initial value of zero) is 
\be
\label{InIn}
\langle \hat x_{\alpha'} \hat x_{\beta'} \hat x_{\gamma'} \rangle = -i  \int^{t}_{t_0} \di t'  \langle \left [\hat x_{\alpha'} \hat x_{\beta'} \hat x_{\gamma'}, {\cal \hat H}_{\rm int}(t') \right ] \rangle\,,
\ee
where ${\cal H}_{\rm int}$ is the interaction Hamiltonian. Here the  interaction picture (as opposed to the Heisenberg picture) is employed. 
Nevertheless, it is 
straightforward to verify that Eq.~\eref{InIn} leads directly to Eq.~\eref{alphaSol2q} using the expressions 
\begin{eqnarray}
\left [\hat x_{\alpha'} \hat x_{\beta'} \hat x_{\gamma'}, {\cal \hat H}_{\rm int}(t') \right ] &=& \Gamma_{  \alpha' \tilde \nu'} \left [ \hat x_{\tilde \nu'}, {\cal \hat H}_{\rm int}(t')\right ] \hat x_{\beta'} \hat x_{\gamma'}
				 + \Gamma_{  \beta' \tilde \nu'}\hat x_{\alpha'} \left [\hat x_{\tilde \nu'}, {\cal \hat  H}_{\rm int}(t') \right ] \hat x_{\gamma'} 
				\nonumber \\ &+& \Gamma_{  \gamma' \tilde \nu'}\hat x_{\alpha'} \hat x_{\beta'} \left [ \hat x_{\tilde \nu'}, {\cal \hat H}_{\rm int}(t') \right ]\,,
\label{InIn2}
\end{eqnarray}
and
\be
\left [\hat x_{\tilde \alpha' }, {\cal H}_{\rm int}(t') \right ]= i  u_{\tilde \alpha' \tilde \nu' \tilde \mu' } \hat x_{\tilde \nu'} \hat x_{\tilde \mu'}\,,
\ee
and employing Wick's theorem. To arrive at Eq.~\eref{InIn2}, we have used the first order part of the expression
\be
\label{taylor}
\hat x_{\alpha'} = \Gamma_{\alpha' i'} \hat x_{i'} + \frac{1}{2}\Gamma_{\alpha' i'j'} \left( \hat x_{i'} \hat x_{j'} - \langle \hat x_{i'} \hat x_{j'} \rangle \right)\,,
\ee
which is the quantum version of the result that the 
$\Gamma$ functions can be identified with the coefficients of a 
$\delta N$ style Taylor expansion, see Refs.~\cite{Seery:2012vj,Anderson:2012em} for a full 
discussion of the classical result. The result follows from substituting Eq.~\eref{taylor} into the right and left hand side of Eq.~\eref{operatorPerts101}, and 
comparing with Eqs.~\eref{gamma1Evolve} and \eref{gamma2Evolve}.

\subsection{Symmetrised correlation functions}

For the equations of motion for the correlation functions themselves, 
a simplification occurs if we choose to deal with fully symmetrised (Weyl ordered) 
correlation functions. Labelling 
\be
\Sigma_{(\alpha' \beta')} \equiv \Sigma^{w}_{\alpha' \beta'}\,,
\ee
and similarly the fully symmetric ordered three-point function as $\alpha^{ w}_{\alpha' \beta' \gamma'}$, we 
find
\begin{eqnarray}
\label{tweyl}
\frac{\di\Sigma^{w}_{\alpha^{\prime}\beta^{\prime}}}{\di t}
	&=&
		u_{\alpha^{\prime}\gamma^{\prime}}
		\Sigma^w_{\gamma^{\prime}\beta^{\prime}}
		+
		u_{\beta^{\prime}\gamma^{\prime}}
		\Sigma^w_{\gamma^{\prime}\alpha^{\prime}}\,, \nonumber \\
\frac{\di\alpha^w_{\alpha^{\prime}\beta^{\prime}\gamma^{\prime}}}{\di t}
	&=&
		u_{\alpha^{\prime}\lambda^{\prime}}
		\alpha^w_{\lambda^{\prime}\beta^{\prime}\gamma^{\prime}}
		+u_{\alpha^{\prime}\lambda^{\prime}\mu^{\prime}}
		\Sigma^w_{\lambda^{\prime}\beta^{\prime}}
		\Sigma^w_{\mu^{\prime}\gamma^{\prime}} - \frac{1}{3} u_{\alpha^{\prime}\lambda^{\prime}\mu^{\prime}}
		\Sigma^i_{\lambda^{\prime}\beta^{\prime}}
		\Sigma^i_{\mu^{\prime}\gamma^{\prime}}\\ \nonumber
	         & + &\left (		\alpha'\rightarrow\beta'\rightarrow\gamma' \right)\,,
\end{eqnarray}
where $\Sigma^i_{\lambda^{\prime}\beta^{\prime}} \equiv -i \Sigma_{[\lambda\beta]} $ is the imaginary part of the two point function, whose 
real part is the Weyl ordered two point function $\Sigma^w_{\gamma^{\prime}\beta^{\prime}}$. 
A possible advantage of writing the system in this way is that we deal only with real numbers.
It is clear that as the dynamics become classical and the imaginary part of the two point function tends to zero, Eqs.~\eref{tweyl} tend to Eqs.~\eref{tbasic}, and hence the classical transport equations are recovered 
from the quantum ones.

\section{Single field inflation}
\label{singleField}
As a simple example of the framework we have been 
discussing, we now consider single field inflation explicitly. 
We follow closely the study of Seery, Malik and Lyth \cite{Seery:2008qj}. In that study the authors showed how 
to calculate non-Gaussianity directly from the the field equations using a Greens function approach. 
The transport approach offers an alternative approach which also follows from the field equations. 
In order to keep things simple, the non-linear part of the field equations are truncated at leading order in slow-roll and we 
employ conformal time, such that $\tau \to -\infty$ is the infinite past and $\tau \to 0$ the infinite future. 
We choose the basic variables 
to be $\delta \phi$ and $\delta  \phi' = \di \delta \phi / \di \tau$. Since there are two degrees of freedom the subscript 
indices run from $1$ to $2$, which in an abuse of notion 
we label $\phi$ and $p$  respectively (even though $\delta \phi'$ is not in general the canonical momentum).
In the  language of the transport formalism we have been developing, we have 
\begin{eqnarray}
u_{\phi p}(k) &=&1 ,\nonumber \\
u_{p \phi}(k) &=& - k^2,\nonumber \\
u_{p p}(k) &=&  -2 {\mathcal H},
\label{u1}
\end{eqnarray}
and
\begin{eqnarray}
u_{p \phi \phi}(k_1,k_2,k_3)&=& - a^2 V'''(\phi) - \frac{\phi'}{\mathcal H} \left(
			 \frac{3}{2} (k_2^2+k_3^2)-\frac{(k_2^2-k_3^2)}{4 k_1^2} \frac{k_1^2}{4} \right )\,,\nonumber\\
u_{p p p}(k_1,k_2,k_3) &=&  - \frac{\phi'}{\mathcal H} \frac{(k_2^2+k_3^2)(k_2^2+k_3^2-k_1^2)}{2 k_2^2 k_3^2}\,,
\label{u2}
\end{eqnarray}
where ${\cal H}= (\di a/\di \tau)/a$. Other components of the $u$ coefficients are zero 
(at least to first order in slow-roll). These 
equations constitute Eqs.~(15)-(16) of \cite{Seery:2008qj} (see also Refs.~\cite{Malik:2007nd} and \cite{Malik:2006ir}).

Our choice of variables is the simplest choice possible, and for example implies that 
$\Gamma_{p i} = \di{\Gamma}_{\phi i}/\di \tau$ and $\Gamma_{p i j} = \di \Gamma_{\phi i j}/\di \tau$, 
and so they both satisfy second order (in time) differential equations. $\Gamma_{\phi i}$, for example 
is a solution to the linear part of the second order 
field equation for $\delta \phi$. 
Given that it obeys  the linear field equation and has the initial 
conditions discussed in \S\ref{transportProperties}, namely that 
$\Gamma^{\rm init}_{\phi p} = 0$ and $\dot \Gamma^{\rm init}_{\phi p} = \Gamma^{\rm init}_{p p} = 1$, 
we can immediately identify $\Gamma_{\phi p}$ as the Greens function 
denoted by Seery {\it et al.} \cite{Seery:2008qj} as $Gr_k$. 

By solving the relevant equation of motion we find, in the de-Sitter 
approximation for which $a \propto (-\tau)^{-1}$, 
\be
\label{Gamma}
\Gamma_{\phi p}(k, \tau, \tau') = \frac{k(\tau' - \tau) { \cos}\left [k(\tau' - \tau)\right ] - (1+k^2 \tau' \tau) {\sin}\left[k(\tau' - \tau)\right ]}{k^3 \tau'^2}\,,
\ee
where, $\tau$ is the later time, and $\tau'$ the earlier time (which we will soon use as an intermediate time in an integral solution).
This expression can be shown to be in agreement with Eq.~(12) of Ref.~\cite{Seery:2008qj}.

Using the formal integral solution Eq.~\eref{alphaSolq}, the three-point can be calculated. Doing so we find identical expressions to those of Seery {\it et al.}. The solution can broken into 
pieces for each non-linear term in the original field equation. Considering the $u_{p \phi \phi}$ terms one finds
\begin{eqnarray}
\alpha_{\phi \phi \phi}(k_1,k_2,k_3) &\supseteq& \int^{\tau}_{-\infty} d \tau' \left [ \Gamma_{\phi p}(k_1, \tau, \tau') u_{p \phi \phi}(k_1,k_2,k_3) \Sigma_{\phi \phi}(k_2,\tau', \tau) \Sigma_{\phi \phi}(k_3, \tau',\tau)\right.\nonumber \\ 
			  &+& \Gamma_{\phi p}(k_2,\tau,\tau') u_{p \phi \phi}(k_2,k_1,k_3) \Sigma_{\phi \phi}(k_1,\tau, \tau') \Sigma_{\phi \phi}(k_3,\tau',\tau)\nonumber \\
	 	 	& +&  \left. \Gamma_{\phi p}(k_3,\tau,\tau') u_{p \phi \phi}(k_3,k_1,k_2) \Sigma_{\phi \phi}(k_1,\tau,\tau') \Sigma_{\phi \phi}(k_2,\tau,\tau') \right ]\,,
\end{eqnarray}
where the two point function is to be 
approximated by the de-Sitter expression, and we have
\begin{eqnarray}
\langle \delta \phi(\mathbf{k},\tau) \delta \phi(\mathbf{k_2},\tau')\rangle &=& \Sigma_{\phi \phi}(k, \tau, \tau')\delta(\mathbf{k}+\mathbf{k_2})\,, \nonumber \\
\Sigma_{\phi \phi}(k, \tau, \tau') &=&  \frac{H^2}{2 k^3} e^{-i k( \tau - \tau')} (1+ i k \tau) (1- i k \tau')\,.
\label{S1}
\end{eqnarray}
For example, considering the potential term in isolation ($u_{p \phi \phi} \supseteq- a^2 V'''(\phi)$),  we find an integral of the form of  
Eq.~(18) of Ref.~\cite{Seery:2008qj}, and performing the integration\footnote{Note the the integral must be 
regulated in the usual way to select the correct vacuum \cite{Maldacena:2002vr}.} one finds 
\be
\alpha_{\phi \phi \phi}(k_3,k_1,k_2) \supseteq
		\frac{H_\ast^2 V'''_\ast}{4 \prod_i k_i^3}   \left(
			- \frac{4}{9} k_t^3 + k_t \sum_{i < j} k_i k_j +
			\frac{1}{3} \Big\{ \frac{1}{3} + \gamma +
				\ln | k_t \eta_\ast | \Big\} \sum_i k_i^3
		\right) \,,
	\ee
	where $*$ indicates that background 
quantities are approximated by fixing them to the value they have when the scales of 
interest crossed the horizon,  $i \in \{ 1, 2, 3 \}$,  $k_t = \sum_i k_i$, and $\gamma$ is Euler's constant.

Using Eqs.~\eref{u2}, \eref{Gamma} and \eref{S1},
together with
\be
\Sigma_{ \phi p}(k,\tau,\tau') = \frac{H^2}{2 k^3} e^{-i k( \tau - \tau')} (1+ i k \tau) \tau'\,,
\ee
the relevant integrals for the remaining parts of $u_{p \phi \phi}$ and the $u_{p p p}$ terms can be shown to be 
of the form of Eqs.~(21) and (25) of Ref.~\cite{Seery:2008qj}.
 Combining these terms one finds 
\be
		\alpha_{\phi \phi \phi}(k_3,k_1,k_2) \supseteq 
\frac{H_\ast^4}{8 \prod_i k_i^3} \frac{\dot{\phi}_\ast}{H_\ast}
		\left\{
			\frac{1}{2} \sum_i k_i^3 -
			\frac{4}{k_t} \sum_{i < j} k_i^2 k_j^2 -
			\frac{1}{2} \sum_{i \neq j} k_i k_j^2
		\right\} .
	\ee
This expression is 
in agreement with Eq.~(29) of Ref.~\cite{Seery:2008qj} and results found using the In-In formalism \cite{Maldacena:2002vr, Seery:2005gb}. 
	In this section we have used the analytic solutions which follow from the transport equations to find the 
three-point function of field fluctuations in single field inflation.This was to show directly in this simple example that 
the transport framework matches others in the literature. We could of course also have solved directly the coupled
ODEs for the two- and three-point functions, Eqs.~\eref{tqbasic}, using a numerical implementation of the 
equations. This would enable us to drop the slow-roll and de-Sitter approximations, and is an issue we will return to in 
future work.

\section{From field perturbations to $\zeta$}

It is often the case that the variables we choose to calculate with are not those which we finally 
want the correlations of. This is the case above where we calculate the three point function of the field 
fluctuation, but ultimately require the correlation functions of $\zeta$.
We now briefly describe how this is accommodated within our framework. 

Classically, $\zeta$ can be written in terms of any complete set of perturbations in the general form 
\be
\zeta(k) = N_{\alpha'} x_{\alpha'} + \frac{1}{2}N_{\alpha' \beta'} \left( x_{\alpha'} x_{\beta'}-\langle x_{\alpha'} x_{\beta'} \rangle \right)\,,
\ee
were $N_{\alpha'}=(2 \pi)^3 N_\alpha(k)\delta(\mathbf{k} - \mathbf{k_\alpha})$ 
and $N_{\alpha' \beta'}= (2\pi)^3 N_{\alpha \beta}(k, k_\alpha, k_\beta)\delta(\mathbf{k} -\mathbf{k_\alpha} - \mathbf{k_\beta})$.
We stress these are not the $\delta N$ coefficients used in the $\delta N$ expansion, but are simply a compact way in 
which to write how $\zeta$ at a given time is related to the fluctuations $x_{\alpha'}$ at the same time. 
In the classical superhorizon regime we find
\be
\langle \zeta \zeta \rangle = N_{\alpha'} N_{\beta'} \Sigma_{\alpha' \beta'}\,,
\ee
and 
\begin{eqnarray}
\label{zetaC}
\langle \zeta \zeta \zeta \rangle &=& N_{\alpha'} N_{\beta'} N_{\gamma'} \alpha_{\alpha' \beta' \gamma'} + N_{\alpha' \beta'} N_{\gamma'} N_{\tau'} \Sigma_{\alpha' \gamma'}\Sigma_{\beta' \tau'}\,.  
\end{eqnarray}
As in the previous sections, the presence of a delta function in both the definitions of the $N$ functions and the 
correlations of $x_\alpha'$ implies 
that the integrations over Fourier space can be performed immediately leaving an expression with no convolutions present.
We note that we are free to perform such a conversion to 
$\zeta$ on large scales where the kernels of the $N$ quantities lose any Fourier dependence. 

The method of calculation is given elsewhere \cite{Anderson:2012em}, but here we simply note that using separate-universe techniques, the $N$ functions can be calculated 
for any cosmology in which the energy density is a function of the unperturbed quantities the $x_\alpha$ variables are perturbations of. Denoting these 
unperturbed quantities $X_\alpha$, so that for example if $x_1= \delta \phi_1$, $X_1 =\phi_1$, one finds
\be
N_\alpha=-\frac{1}{2 \dot{H}} \frac{\partial H^2}{ \partial X_\alpha}\,,
\ee
and 
\be
N_{\alpha \beta}=-\frac{1}{2 \dot{H}} \frac{\partial^2 H^2}{ \partial X_\alpha X_\beta} - \frac{\partial}{ \partial X_{(\alpha}}\left(\frac{1}{\dot{H}}\right)
\frac{\partial H^2}{ \partial X_{\beta)}}+\frac{1}{2 \dot{H}}\frac{\partial}{\partial X_\gamma}\left(\frac{1}{2 \dot{H}}\right) \frac{\di X_\gamma}{\di N} \frac{\partial H^2}{\partial X_\alpha} \frac{\partial H^2}{X_\beta}\,,
\ee
where $H$ and $\dot H$ are the Hubble rate and its derivative with respect to cosmic time, and are to be treated as functions of the underlying variables $x_\alpha$.
\subsection{Single field inflation}
Returning to the case of single field inflation under the slow-roll assumption, we have
\begin{eqnarray}
H^2 &=& \frac{1}{3\Mpl^2} V(\phi)\,,\nonumber \\
\dot{H} &=& \frac{-V_{,\phi} V_{,\phi}}{6 V} ,
\end{eqnarray}
and employing the slow-roll parameters 
\begin{eqnarray}
\epsilon = \frac{\Mpl^2}{2} \left( \frac{V_{,\phi}}{V} \right )^2\,,~~~ \eta = \Mpl^2\frac{V_{,\phi\phi}}{V}\,,
\end{eqnarray}
one finds
\begin{eqnarray}
N_\phi &=& \frac{1}{\sqrt{2 \epsilon}}\,,\nonumber \\
N_{\phi \phi} &=& \left ( 1 - \frac{\eta}{2 \epsilon}   \right ) \,.
\end{eqnarray}
These expression are in agreement with those in the literature and in particular with Eq.~(32) of Ref.~\cite{Seery:2008qj}. Using them, 
the bispectrum of $\zeta$ can be calculated using Eq.~\eref{zetaC}, as was done in Ref.~\cite{Seery:2008qj}, once 
again one finds agreement with the results of the In-In formalism as one must.

\section{The power spectrum in multi-field models}
\label{multiField}
Having laid out the sub-horizon transport formalism, and shown how it connects to both the In-In formalism and the field equation 
approach of Seery, Malik and Lyth \cite{Seery:2008qj}, we wish to make one further connection to work in the literature before concluding. 
There is a well established methodology for calculating the 
power spectrum in multi-field models of inflation which includes the sub-horizon evolution. This was originally laid out in 
detail by Salopek, Bond and Bardeen  in Ref. \cite{Salopek:1988qh}, and continues to be applied in numerical studies, for example in recent studies utilising the Pyflation
numerical package \cite{Huston:2011fr} (see also \cite{McAllister:2012am} and references therein for related recent numerical implementations). 
It reflects the simple application of linear quantum field theory to multi-field inflation, and as such 
should agree with and recover the transport equations for $\Sigma_{\alpha \beta}$. We now show that this is indeed the case.

The formalism of Bardeen {\it et al.} begins with the definition for the field fluctuation operators in terms of creation and annihilation operators, 
employing variables $\delta \phi_a$ and $\delta \dot \phi_a$,
we have
\be
\delta \phi_a(t, \mathbf{k}) = \Psi_{a c}(t, k) a_c(\mathbf{k})+  \Psi^*_{a c}(t, k) a^\dagger_c(-\mathbf{k}),
\ee
where $a_c$ and $a^\dagger_c$ are the usual creation and annihilation operators, 
the matrix $\Psi_{a b}$ allows for coupling between the fields, and in this section only, Roman numerals label 
the fields and run from $1$ to $n$ where $n$ is the number of fields.

Once again abusing notation 
by labelling the subscript indices for our $u$ coefficients by $\phi_a$ and $p_a$,  
one can show that the evolution equation for the 
$\Psi_{a b}$ matrix is given by 
\be
\label{bevolve}
\ddot{\Psi}_{ab}(t, k) =  u_{p_a  \phi_c }(k) \Psi_{c b}(t, k) +   u_{p_a  p_c}(k) \dot  \Psi_{c b}(t, k)\,, 
\ee
where we sum over $c$ from $1$ to $n$.

The kernel of the quantum two point function is given by 
\begin{eqnarray}
\Sigma_{\phi_a \phi_b}(t, k)&=& \Psi_{a c}(t, k)\Psi^*_{b c}(t, k),\nonumber \\
\Sigma_{\phi_a \p_b}(t, k)&=& \Psi_{a c}(t, k)\dot \Psi^*_{b c}(t, k),\nonumber \\
\Sigma_{\p_a \p_b}(t, k)&=& \dot \Psi_{a c}(t, k)\dot \Psi^*_{b c}(t, k).
\end{eqnarray}
Employing these relations, together with the evolution 
equation \eref{bevolve}, to find an evolution equation for $\Sigma_{\alpha \beta}(k)$, 
and recalling that $u_{\phi_a p_b}(k) = \delta_{a b}$ and $u_{\phi_a \phi_b}(k) =0$
we readily find that we arrive back at the equations of motion for $\Sigma_{\alpha \beta}(k)$ we 
found in \S\ref{quantumTransport}. This can be viewed as a consistency check, or an alternative 
derivation of the transport equations for the quantum two-point function.

\section{Conclusions}
\label{conclusions}
In this work we have shown how the transport equations for the correlation functions of inflationary perturbations
can be extended from purely classical super-horizon equations to equations which are also valid on 
sub-horizon scales. This provides ODEs for the quantum statistical properties of perturbations 
valid on all scales. A property of these equations is that they do not involve a convolution over all scales, even for non-linear 
statistics, unlike the non-linear field equation for a given Fourier mode of a perturbation. In this sense evolution equations for statistics are simpler than 
those for the perturbations themselves. An alternative system of coupled ODEs which also exhibits this feature, and allows the 
calculation of inflationary statistics, is that of the $\Gamma$ objects, introduced in \S\ref{transportProperties}

A major aim of this paper has been to show how this formulation for calculating 
observables form inflation is connected to previous work. In particular we have shown 
how the quantum transport equations reduce to the classical moment transport equations for inflation which have 
been studied in detail in previous work \cite{Mulryne:2010rp,Seery:2012vj,Anderson:2012em}.  
Moreover, we have shown that integral solutions to the transport equations 
are equivalent to the integral equations which arise in the In-In formalism, and identical to the integral Greens function 
solutions found by Seery, Malik and Lyth in the inflationary context \cite{Seery:2008qj}. Finally, we have shown the equivalence to the 
method for calculating the power spectrum in multi-field inflation developed by Salopeck {\it et al.}. Given the previous transport 
studies have established the equivalence of the classical 
transport equations to the $\delta N$ formalism, this new work shows the connections between all methods 
in common use for calculating inflationary observables, both quantum and classical.

The advantage of the transport route to 
non-Gaussianity calculations, however, is the 
presence, and straightforward origin, of the ODEs for quantum correlation functions.
Ultimately we hope that these differential equations will lead to an 
efficient and easy to implement algorithm for the numerical computation of observable 
statistics of perturbations produced by inflation, accounting for the evolution from quantum vacuum to 
super-horizon scales. Numerical studies are necessary for models with non-linear 
terms which cannot be well approximated within the slow-roll framework (for example \cite{Chen:2006xjb,Chen:2008wn}), 
even by studies which go beyond leading order
such as \cite{Noller:2011hd,Burrage:2011hd,Ribeiro:2012ar}, as well as for general multi-field models.
While some numerical work in the direction has already been implemented using alternative 
approaches (for example \cite{Huston:2011vt,Chen:2006xjb,Chen:2008wn,Hazra:2012yn,Funakoshi:2012ms}), it 
stops short of code for the full bispectrum in multiple field models. This is a pressing problem given the 
immense number of inflationary models, and the large number of parameter choices and initial conditions  
possible in each. Without such tools it is impossible to make a judgement about the viability of a 
particular model in the light of present and future data. We will therefore present a numerical 
implementation of our scheme in future work \cite{DFMS}.

\section*{Acknowledgements}
The author is supported by the Science and Technology Facilities Council grant ST/J001546/1. He expresses gratitude 
to Jonathan Frazer and Mafalda Dias for helpful discussions, and to Joseph Elliston, Karim Malik, David Seery and Reza Tavakol for helpful discussions and for careful reading and comments on a draft copy.

\bibliographystyle{JHEP}
\bibliography{paper}

\end{document}